\documentclass[iop,apjl]{emulateapj}
\usepackage{amsmath,apjfonts}

\slugcomment{Accepted ApJL, 2011 Apr 21}
\shorttitle{2010 May Cyg X-3 Flare}
\shortauthors{Williams {\it et al.}}

\newcommand\citeeg[1]{\citep[\textit{e.g.}][]{#1}}
\newcommand\ojd[1]{OJD~#1}
\newcommand\g{\ensuremath{\gamma}}
\newcommand\gr{\g-ray}
\newcommand\grs{\g-rays}
\newcommand\gre{\g-ray emission}
\newcommand\fermi{{\it Fermi}}
\newcommand\integral{{\it INTEGRAL}}
\newcommand\agile{{\it AGILE}}
\newcommand\cyg{\object{Cyg~X-3}}

\newcommand\figslowlc{f1}
\newcommand\figrapidlc{f2}
\newcommand\figrpostflare{f3}

\begin{document}

\title{The 2010 May Flaring Episode of Cygnus X-3 in Radio, X-Rays,
  and \g-Rays}
\author{
  Peter~K.~G.~Williams\altaffilmark{1},
  John~A.~Tomsick\altaffilmark{2},
  Arash~Bodaghee\altaffilmark{2},
  Geoffrey~C.~Bower\altaffilmark{1},
  Guy~G.~Pooley\altaffilmark{3},
  Katja~Pottschmidt\altaffilmark{4,5},
  J\'er\^ome~Rodriguez\altaffilmark{6},
  J\"orn~Wilms\altaffilmark{7},
  Simone~Migliari\altaffilmark{8},
  Sergei~A.~Trushkin\altaffilmark{9}
}

\altaffiltext{1}{Department of Astronomy, 601 Campbell Hall \# 3411,
  University of California, Berkeley, CA 94720-3411, USA;
  \texttt{pwilliams@astro.berkeley.edu}}
\altaffiltext{2}{Space Sciences Laboratory, 7 Gauss Way, University of
  California, Berkeley, CA 94720-7450, USA}
\altaffiltext{3}{The University of Cambridge, Mullard Radio Astronomy
  Observatory, Cavendish Laboratory, J.~J.~Thomson Avenue, Cambridge
  CB3~0HE, UK}
\altaffiltext{4}{CRESST and NASA-Goddard Space Flight Center,
  Astrophysics Science Division, Astroparticle Physics Laboratory,
  Greenbelt, MD 20771, USA}
\altaffiltext{5}{CSST, University of Maryland Baltimore County, 1000
  Hilltop Circle, Baltimore, MD 21250, USA}
\altaffiltext{6}{AIM-Astrophysique Instrumentation Mod\'elisation (UMR
  7158 CEA/CNRS/Universit\'e Paris 7 Denis Diderot), CEA Saclay,
  DSM/IRFU/Service d'Astrophysique, B\^at. 709, L'Orme des Merisiers,
  FR-91 191 Gif-sur-Yvette Cedex, France}
\altaffiltext{7}{Dr. Karl Remeis-Sternwarte and Erlangen Center for
  Astroparticle Physics, Universit\"at Erlangen-N\"urnberg,
  Sternwartstr. 7, 96049 Bamberg, Germany}
\altaffiltext{8}{Departament d'Astronomia i Meteorologia (DAM) and
  Institut de Ci\`ences del Cosmos (ICC), Universitat de Barcelona,
  Mart\'i i Franqu\`es 1, 08028 Barcelona, Spain}
\altaffiltext{9}{Special Astrophysical Observatory RAS,
  Karachaevo-Cherkassian res, Nizhnij Arkhyz 36916, Russia}

\begin{abstract}
  In 2009, \object{Cygnus~X-3} (Cyg~X-3) became the first microquasar
  to be detected in the GeV \gr\ regime, via the satellites \fermi\ and
  \agile. The addition of this new band to the observational toolbox
  holds promise for building a more detailed understanding of the
  relativistic jets of this and other systems. We present a rich
  dataset of radio, hard and soft X-ray, and \gr\ observations of
  \cyg\ made during a flaring episode in 2010~May. We detect a
  $\sim$3-d softening and recovery of the X-ray emission, followed
  almost immediately by a $\sim$1-Jy radio flare at 15~GHz, followed
  by a 4.3$\sigma$ \gr\ flare (E $>$ 100~MeV) $\sim$1.5~d later. The
  radio sampling is sparse, but we use archival data to argue that it
  is unlikely the \gr\ flare was followed by any significant
  unobserved radio flares. In this case, the sequencing of the
  observed events is difficult to explain in a model in which the
  \gre\ is due to inverse Compton scattering of the companion star's
  radiation field. Our observations suggest that other mechanisms may
  also be responsible for \gre\ from \cyg.
\end{abstract}

\keywords{black hole physics --- X-rays: binaries --- X-rays:
  individual (Cygnus~X-3)}

\section{Introduction}
\label{s:intro}

The X-ray binary (XRB) system \object{Cygnus~X-3} (hereafter \cyg),
discovered by \citet{gggw67}, is notable for its Wolf-Rayet (WR)
companion \citep{vk+96}, short (4.8~h) orbital period \citep{bbc88},
and dramatic radio variability \citep{wfjg94}. The relativistic
jets that it produces \citep{gjs+83} classify the source as a
``microquasar'' system. The distance to \cyg\ is $\sim$7 or
$\sim$9~kpc, depending on the method used \citep{pbpt00,lzt09}. It is
also uncertain whether the compact object is a black hole
\citeeg{cm94} or neutron star \citeeg{ss03}.

\cyg\ became more notable in 2009 when it became the first microquasar
to be detected in the GeV \gr\ regime, via the satellites
\fermi\ \citep{flc09} and \agile\ \citep{t+09}. A 4.8-h modulation in
the \fermi\ data makes the detection definitive. Emission of \grs\ has
been detected by both observatories during every ``high/soft'' X-ray
state to occur since they began science operations (\citealp{flc09};
\citealp{t+09}; \citealp{ch10}; \S\ref{s:results}). The high/soft
state is also associated with radio flaring and is closely tied to the
launching of relativistic jets \citeeg{fbg04,khm+10}.

Various physical processes are expected to be involved in the emission
and propagation of high-energy photons in \cyg. Inverse Compton (IC)
upscattering of the intense ($\sim$10$^5$ erg cm$^{-3}$) radiation
field of the WR companion by high-energy ($\g \gtrsim 10^3$) electrons
in the relativistic jet is the most efficient source of GeV emission
\citep{flc09,dch10}. If the density of relativistic protons in the jet
is sufficiently high, inelastic collisions between these and protons
in the stellar wind will produce neutral pions that decay into
\grs\ (among other products) as well \citep{rtkbm03}. In this case,
there will be further \gre\ due to secondary leptons created during
the hadronic interactions \citep{obbr+07}. The typical hadronic
content of microquasar jets is unknown: large numbers of hadrons
impose significant energetic constraints, but there is direct evidence
of their presence in at least one system, \object{SS~433}
\citep{kka+94,kkmb96,mfm02}.

A \gr\ flare due to IC upscattering by jet electrons will be followed
by a radio flare a few days later as those electrons cool, rarify, and
become transparent to their radio synchrotron emission. This
phenomenon has been observed repeatedly but not invariably
\citep{flc09,t+09}. If the \gre\ is primarily hadronic, the result
will be substantially the same, as secondary processes will create a
similar population of energetic leptons. \citet{flc09} determine an
approximate $5 \pm 7$~d lag between radio and \gre. Inspection of
their Fig.~2 and the data presented by \citet{t+09} suggests that this
uncertainty is primarily systematic rather than statistical; {\it
  i.e.}, that the radio/\gr\ lag varies significantly from event to
event. Blazar observations often show lag variability, both between
sources and in the same source over time, with theoretical support for
sign changes in the lag between \gr\ and other bands \citep{aha10}.

In 2010~May, multiple monitoring programs detected that \cyg\ was
becoming more active in \grs\ and was entering the high/soft state
\citep{bss+10,cflc10,kms+10,ch10}. We describe multiband
observations of \cyg\ made during this episode (\S\ref{s:obs}) and
present our results (\S\ref{s:results}).  Our data show a \gr\ flare
apparently lagging a radio flare, a sequencing opposite of that
expected in the typical interpretation. We discuss and interpret this
result (\S\ref{s:discussion}) and, finally, present our conclusions
(\S\ref{s:conc}).

\section{Observations \& Data Analysis}
\label{s:obs}

We observed \cyg\ during its 2010~May flaring episode with a variety
of instruments. Below, we describe the observations and their
analysis. Many dates and times we mention are near MJD 55,340. For
brevity, we express these in an offset MJD, defining $\textrm{OJD} =
\textrm{MJD} - 55,300$. Thus \ojd{0} is JD~2,455,300.5 or
2010~Apr~14.0~UT, and the \gr\ flare was detected on \ojd{43} =
2010~May~27~UT.

\subsection{Radio}
\label{s:radioobs}

The Arcminute Microkelvin Imager (AMI) arrays (Cambridge, UK) are two
aperture synthesis telescopes mainly used to study the cosmic
microwave background \citep{zbb+08}. The observations described herein
were made with the Large Array (AMI-LA), the reconfigured and
reequipped Ryle Telescope, consisting of eight 13-m antennas with a
maximum baseline of about 120~m, observing in the band 12--18~GHz. The
angular resolution is typically 25\arcsec. Monitoring of
small-diameter sources is undertaken as described in \citet{pf97};
observations are interleaved with those of a phase-reference
calibrator, and after calibration the data for individual baselines
are vector-averaged. The in-phase component then provides an unbiased
estimate of the target source's flux density. The amplitude scale was
calibrated by (at least) daily observations of 3C\,48 and 3C\,286,
both of which are believed to be very nearly constant on long
timescales.

Should there be emission on a scale resolved by some of the baselines,
the flux density estimate would be incorrect. There is an extended
region of low-brightness emission around \cyg\ \citep{ssmc+08} which
is detected on the shortest baselines at extreme hour angles with the
AMI-LA, and is sufficiently bright to need separate treatment when the
central source in \cyg\ is ``quenched'' (flux density $\lesssim$
10~mJy at 15~GHz). No such correction was required in the observations
described here.

Observations of \cyg\ were also made with the RATAN-600 telescope
(Nizhnij Arkhyz, Russia) as part of an ongoing microquasar monitoring
campaign \citep{t00,tbn+06}. The observations and data analysis were
performed as described in \citet{tbn+06}. In this work we show results
at 11.2~GHz; near-simultaneous observations were made at 4.8 and
7.7~GHz as well.

\cyg\ was observed by the Allen Telescope Array \citep[ATA; Northern
  California, USA;][]{theata} six times in the period \ojd{26--43} as
part of a larger transient search. Continuum images at 3.09~GHz were
made after subtraction of model of the static sky. Source fluxes were
determined in $\sim$10-minute segments by fitting point source models
in the image domain.

\subsection{X-Ray}
\label{s:xrayobs}

Our study includes hard X-ray observations made with the {\em
  International Gamma-Ray Astrophysics Laboratory} (\integral)
satellite \citep{winkler03}. We used the \integral\ Soft Gamma-Ray
Imager \citep[ISGRI;][]{theisgri} instrument and included two
observations that were part of a program to observe \object{Cyg~X-1}
(PI: Wilms). Although the primary target of the observations was
\object{Cyg~X-1}, the ISGRI FOV is large enough to include the entire
Cygnus region, and our target was in the FOV throughout the
observations. The first observation occurred during revolution 929,
starting at \ojd{40.625} and ending at \ojd{41.940}. The second
observation occurred during revolution 938, starting at \ojd{67.469}
and ending at \ojd{68.805}. A preliminary report of the first
observation was given in \citet{ttf+10}.

We reduced the ISGRI data using the Off-line Scientific Analysis (OSA
v9.0) software package. The program \texttt{ibis\_science\_analysis}
is the primary tool for extracting the standard data products. We
produced and inspected the ISGRI image in the 20--40~keV band, and
then made energy spectra and light curves for all of the bright
sources in the field. After correcting for instrumental deadtime, the
total exposure times for \cyg\ from the first and second observations
are 62,420~s and 66,540~s, respectively. We produced 20--40~keV light
curves with a time resolution of 1~ks.

We also obtained one-day average quick-look measurements from the
All-Sky Monitor on the {\it Rossi X-Ray Timing Explorer} \citep[{\it
    RXTE}/ASM;][]{therxteasm} and transient monitor results from the
Burst Alert Telescope on the {\it Swift} \gr\ burst mission
\citep[{\it Swift}/BAT;][]{theswiftbat}. These results are provided
publicly by the {\it RXTE}/ASM and {\it Swift}/BAT teams,
respectively.

\subsection{\g-Ray}
\label{s:fermiobs}

\fermi\ Science Tools 9.17 and HEASoft 6.9 were used to reduce and
analyze all \fermi-LAT \citep{atwood09} observations within
10\degr\ of \cyg\ that took place in the range \ojd{$-59$--150}
(2010~Feb~14 -- 2010~Sep~11~UT), providing a baseline of $\sim$100~d
before and after the flare of \ojd{43--44}. \cyg\ is within the LAT
FOV for $\sim$15~ks per day. Given this and the typical flux of
\cyg\ in the LAT band, we were compelled to use 1-day time bins in the
following analysis.

Thirty arcminutes away from \cyg\ is a comparatively bright pulsar,
\object{PSR~J2032+4127}. Using \texttt{gtephem} and \texttt{gtpphase},
we added the phase from the most recent ephemeris available
\citep{flc09} to the events file. Then, we extracted events
corresponding to the off-pulse phases \citep[0--0.12, 0.2--0.6, and
  0.72--1;][]{c+09}. This removed 20\% of the live time, so the
exposure time of the off-pulse events file was corrected accordingly.

With \texttt{gtselect} and \texttt{gtmktime}, good time intervals from
the off-pulse events file were selected in the 0.1--10~GeV energy
range. At high energies ($\gtrsim$10~GeV), \cyg\ is expected to have
negligible emission. The events class was set to ``3'' which selects
only high-quality diffuse-background photons. To minimize background
albedo photons from the Earth's limb, zenith and rocking angles were
restricted to less than 105\degr\ and 52\degr, respectively.

Exposure maps were generated by \texttt{gtexpcube} and
\texttt{gtexpmap} while \texttt{gtbin} created photon counts maps in
the region of interest. Emission from \cyg\ was not apparent on the
counts maps, which is unsurprising given that \cyg\ is a faint source
whose peak emission is at the low end of the LAT band where the point
spread function is 5--10\degr. Complicating matters is the high level
of diffuse background emission.

Following the procedure described in \citet{flc09}, unbinned
likelihood analysis was performed for the daily bins with
\texttt{gtlike} considering photons inside a 7\degr\ radius of
\cyg\ from all 1FGL \citep{1fgl} sources up to 5\degr\ away, and all
bright (detection significance $> 7\sigma$ and flux [$>$100~MeV] $>
5\times10^{-8}$~ph cm$^{-2}$ s$^{-1}$) 1FGL sources up to
20\degr\ away. The instrument response function was ``Pass 6 v3''
(\texttt{P6\_V3\_DIFFUSE}), and the convergence relied on the
\texttt{NEWMINUIT} method. Spectral models and input parameters from
the 1FGL catalog were used except for pulsars
\object{1FGL~J2021.0+3651} and \object{1FGL~J2021.5+4026} which had
exponential cutoffs in addition to their power laws. Besides \cyg,
whose photon index and normalization were free, all sources had
spectral parameters fixed to the 1FGL values. We included models for
the galactic (\texttt{gll\_iem\_v02}) and extragalactic
(\texttt{isotropic\_iem\_v02}) diffuse emission. The normalizations
were left free to vary, as was the photon index of the galactic
component.

\section{Results}
\label{s:results}

\begin{figure*}[t]
\plotone{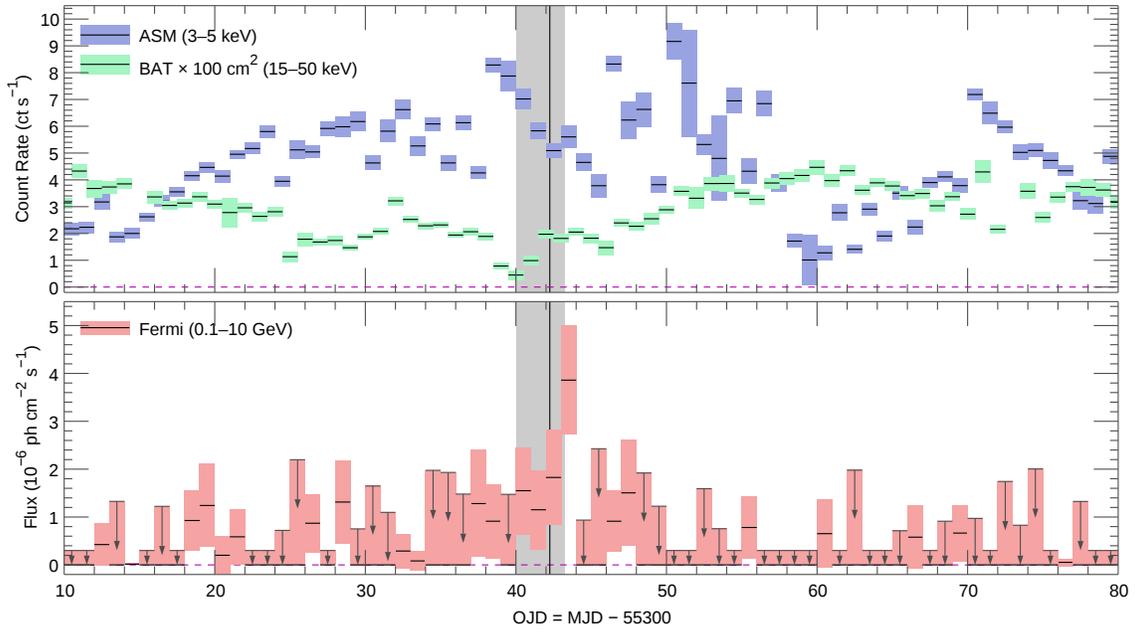}
\caption{Long-term lightcurve of Cyg~X-3 in X- and \grs. {\it
    Shaded region}: timespan depicted in Figure~\ref{f:rapidlc}. {\it
    Vertical black line}: time of maximum observed 15~GHz flux
  density. On some days, our derived {\it Fermi} upper limits are
  smaller than the canonical background rate, $0.3 \times 10^{-6}$ ph
  cm$^{-2}$ s$^{-1}$; on those days, we use the background rate as the
  upper limit.}
\label{f:slowlc}
\end{figure*}

Long-term lightcurves of \cyg\ in soft and hard X-ray bands are
presented in the top panel of Figure~\ref{f:slowlc}. The system was in
a soft X-ray state during \ojd{$\sim$20--55}, with an episode of
particular softness occurring during \ojd{39--41}.

The \fermi\ source flux integrated over 0.1--10~GeV is presented in
the lower panel of Figure~\ref{f:slowlc}. In bins where the test
statistic (TS $\sim -2\ln L$, where $L$ is the ratio of the likelihood
of models without and with the source) is less than one, we compute
and plot a 1$\sigma$ upper limit. A peak is seen starting on
\ojd{43}. The TS value in this bin is 18.4, which translates
approximately to 4.3$\sigma$, and the flux is $(4 \pm 1) \times
10^{-6}$~ph cm$^{-2}$ s$^{-1}$. We are unable to achieve a more
precise timing of the flare. Shifting the binning by half a day
results in lower-significance and lower-flux detections in the range
\ojd{42.5--44.5}. Analysis using higher-cadence binning does not yield
significant detections of the peak.

\begin{figure*}[t]
\plotone{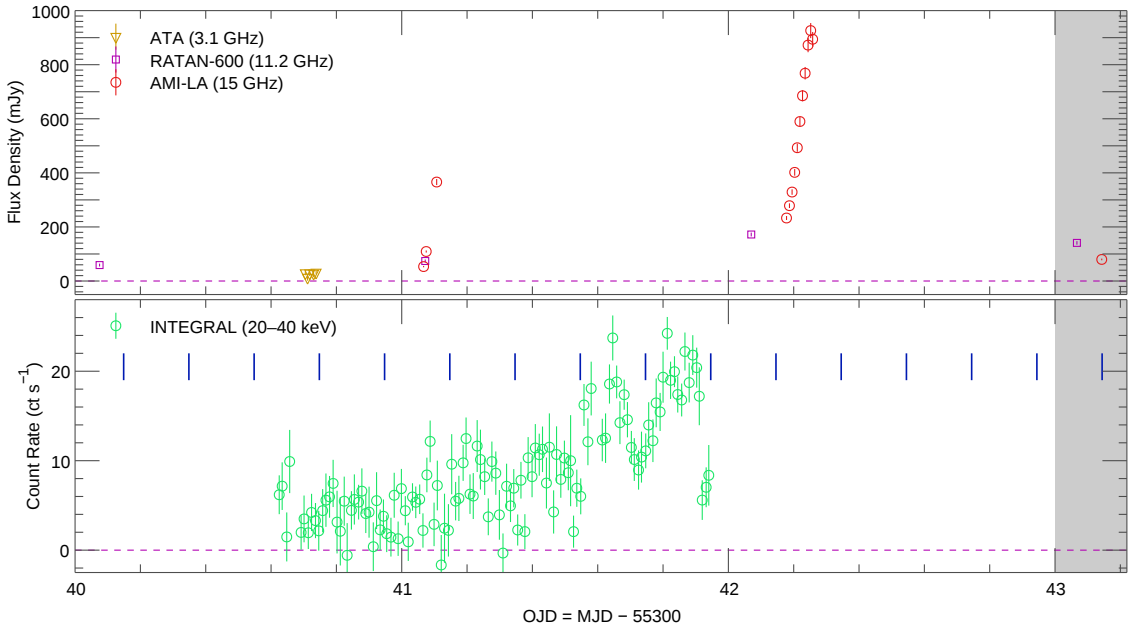}
\caption{High-cadence lightcurve of Cyg~X-3 in radio and hard X-rays
  in the time immediately before the \gr\ peak (= \ojd{43--44}). The
  X-rays are subject to orbital modulation with a 4.8-h periodicity.
  {\it Vertical blue markers}: times of X-ray minima according to
  parabolic ephemeris of \citet{snp+02}. {\it Shaded region}: partial
  timespan of \fermi\ peak.}
\label{f:rapidlc}
\end{figure*}

\begin{figure*}[t]
\plotone{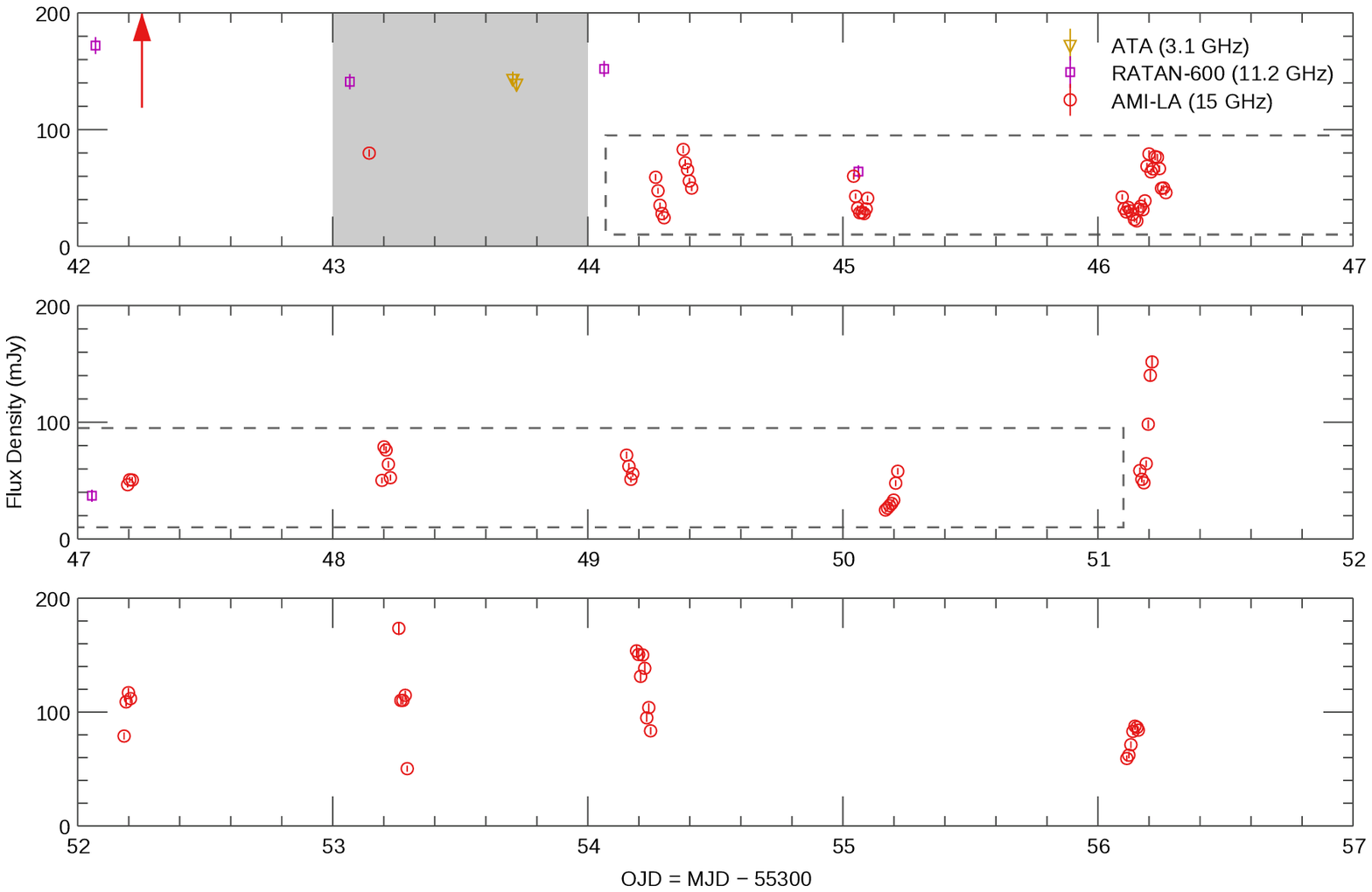}
\caption{Radio lightcurve of Cyg~X-3 during and after the
  \gr\ flare. The three panels proceed in chronological order from top
  to bottom. {\it Red arrow}: time of maximum observed 15~GHz flux
  density (= 926~mJy). {\it Shaded region}: timespan of the
  \fermi\ peak (= $(4 \pm 1) \times 10^{-6}$~ph cm$^{-2}$ s$^{-1}$).
  {\it Dashed box}: period of quiescent radio emission.}
\label{f:rpostflare}
\end{figure*}

Radio measurements from just before the \gr\ flare are presented in
the top panel of Figure~\ref{f:rapidlc}. A $\sim$1~Jy radio flare
occurred on \ojd{42.25}. There is evidence of an earlier flare on
\ojd{41.11}, with a maximum observed flux density of 366~mJy, but only
the rising portion of the flare is observed, so its properties are
poorly-constrained. As is shown in Figure~\ref{f:rpostflare}, the
15~GHz flux density of \cyg\ is depressed in the seven days following
the \gr\ flare, with a mean flux density of 48~mJy and an
80$^\textrm{th}$-percentile observed flux density of 66~mJy. (That is,
80\% of the measurements during this period are $< 66$~mJy.) Starting
on \ojd{51}, the typical radio flux density increases by a factor of
$\sim$2 to $\sim$100~mJy.

The bottom panel of Figure~\ref{f:rapidlc} shows high-cadence hard
X-ray measurements from before the \gr\ flare, made as \cyg\ exited
the softest portion of its soft state. They show the recovery of the
hard X-ray flux, subject to orbital modulation. The modulation makes
it difficult to determine the significance of the drop in flux seen in
the last three measurements; it may be evidence of a significant
reduction in hard X-ray emission precursing the radio flare. Combining
the {\it Swift}/BAT and \integral\ measurements, the hard X-ray flux
has recovered by \ojd{$\sim$42.0}, about 5~h before the observed radio
flare. The \gr\ flare then occurred $\sim$1--2~d later.

\section{Discussion}
\label{s:discussion}

If our observations are taken at face value --- that is, one assumes
no significant radio activity during coverage gaps and that the
flaring stems from one ejection event --- the sequencing of the radio
and \gre\ is inconsistent with the companion IC model discussed in
\S\ref{s:intro}. Emission in \grs\ following a radio flare could be
interpreted as a reenergization of the relativistic jet electrons by a
shock \citep{dch10}. With a typical jet speed of $\sim$0.5~$c$
\citep{tfg+07} and a delay of $\sim$1~d, the reenergization would
occur $\sim$100~AU from the system. This is much larger than the
orbital separation but three orders of magnitude smaller than the
distance at which the jet is expected to interact significantly with
the ISM ($\sim$1~pc). Such reenergization could be due to a collision
between the jet and a dense clump of the WR wind \citep{abrr09}. In
this scenario, the absence of \gre\ at the time of ejection can be
explained by absorption within the system \citep{cdm+11arxiv},
although this model must be reconciled with the observations of
\citet{t+09} and \citet{flc09}. The reenergization would lead to
additional radio emission as the \gr-emitting electrons cool, although
the ambient magnetic field, and hence synchrotron luminosity, will be
much weaker than that found close to the system.

If the \gr\ flare is interpreted as the result of a discrete ejection
event, the lack of a notable subsequent radio flare could be explained
by the ejected material being largely hadronic. While hadronic
\gre\ is not as efficient as IC upscattering, it does yield a higher
ratio of \gr\ to radio luminosity \citep{rv08}. The secondary leptons
due to hadronic interactions would, however, radiate, and detailed
studies typically find that their bolometric luminosity is comparable
to that of the primaries \citeeg{vr10}.

Should our data be taken at face value? Application of a simple
synchrotron-cloud model \citep{vdl66} supports the intuition that
radio flares with sizes comparable to that of the largest observed
could have occurred without detection, with the modeled lifetimes
being $\sim$0.2~d. While this model has had ambiguous success when
applied to \cyg\ \citep{fbw+97}, its simplicity is advantageous for
our sparse data, and it has been successfully applied to observations
of other systems \citeeg{fpbn97,wpp+07,prp10}. To obtain a better
understanding of the radio behavior of \cyg\ we compared the 2010~May
data to 15 years of archived AMI-LA / Ryle Telescope observations. The
$\sim$1~Jy radio flare is unusual: no flux densities $> 400$~mJy are
detected in a 500~d span around the observations we describe.

In light of the predictions of the companion IC model, we consider
particularly the radio observations after the observed \gr\ flare.
The duty cycle of the observations is only 6.8\% in the seven-day
post-flare quiescent interval. We identified 51 epochs similar to this
in the archives, each lasting at least 7 days and having an
80$^\textrm{th}$-percentile observed flux less than 70~mJy
(cf. \S\ref{s:results}). These archival measurements have a duty cycle
comparable to that of the 2010~May data (4.9\%), but a much larger
time on-source (55.5 days). Because this time is large compared to the
total duration of the 2010~May post-\g-flare quiescence, a search for
flares in the archival dataset can constrain the likelihood of there
having been an unobserved flare in the 2010~May dataset, assuming no
long-term evolution in the behavior of the quiescent state and
stochastic flaring.

The epochs were identified by exhaustively searching for seven-day
segments meeting the aforementioned criteria, then lengthening these
segments as much as possible without violating the
80$^\textrm{th}$-percentile constraint. Segments separated by less
than three days were merged, in a few cases shortening the segment
somewhat to preserve the statistical constraint. This method of
construction does not bias against epochs containing rapid flares, no
matter their size.  Qualitatively, the lightcurve of the typical epoch
starts high, drops to very low flux densities, and then becomes high
again, possibly with rapid flares in the middle.

The largest flux density seen in the selected archival data was
524~mJy, in the context of a single rapidly-evolving (risetime
$\sim$0.1~d) flare that would not have necessarily been detected in
the 2010~May observing. (Here we exclude slow, large,
epoch-terminating flares that would have been easily seen in the
2010~May data.) Approximating the rate of such flares as one per 55.5
days of observing, we derive an 11\% chance that such a flare occurred
in the 7 days after the gamma-ray peak without being
observed. Ignoring the epoch beginnings and endings, which contain
high flux values by construction, we find that the flux density of
\cyg\ is larger than 250~mJy only 0.3\% of the time during its
quiescent state.

\section{Summary \& Conclusions}
\label{s:conc}

We have presented observations of a 2010~May \cyg\ flaring episode in
the radio, soft and hard X-ray, and \gr\ bands. Our data show a
sequence of three notable events: a particular softening and partial
recovery of the X-ray emission, a rapid $\sim$1~Jy radio flare, and a
\gr\ flare. Interpretation of the data is made more challenging by the
sparse sampling of the radio data and the low cadence of the
high-energy observations as compared to the rapidity with which the
radio emission can evolve. Nonetheless, the observations we do have
and comparisons to historical data challenge the interpretation that
the \gre\ is due to IC upscattering of the companion radiation field
by high-energy electrons leaving the system in a relativistic jet,
because there is no evidence for the subsequent radio emission that
one would expect to see from these electrons.

While the companion IC model of \gre\ from microquasars is clear and
compelling, the lack of consistent radio/\gr\ timing lags call into
question its completeness. There could be more than one mechanism
responsible for the \gre\ of \cyg, becoming more or less relevant in
different circumstances, or the reprocessing of the \gre\ by effects
such as pair production could be more significant than commonly
assumed. Detections of or limits to very high-energy
\grs\ \citep[$\gtrsim$100~GeV; e.g.][]{magic10} or neutrinos from
\cyg\ would aid in the understanding of the processes relevant to the
emission in the \fermi\ band.

The power of multiband monitoring of \cyg\ promises to increase
significantly with the addition of \gr\ data to the set of available
observations. As \cyg\ inevitably produces more flares, frequent
observations at all wavelengths are important to build a detailed
understanding of the launching and propagation of its relativistic
jets. Radio observations of $\lesssim$2~h cadence with near-continuous
coverage would be ideal for establishing an unambiguous relationship
between radio and \gr\ flaring. Based on the phenomenology we observe,
intensive radio observations should be triggered at the ends of very
soft X-ray states without waiting for the detection of \gr\ flares.

\acknowledgments

The authors thank S.~Corbel for useful discussions. PKGW was supported
by a Space Sciences Lab Summer Fellowship. PKGW, JAT, and AB
acknowledge partial support from Fermi and INTEGRAL Guest Investigator
grants NNX08AW58G, NNX08AX91G, and NNX10AG50G. SM acknowledges
financial support from MEC and European Social Funds through a Ram\'on
y Cajal and the Spanish MICINN through grant
AYA2010-21782-C03-01. Research with the ATA is supported by the Paul
G. Allen Family Foundation, the National Science Foundation, the US
Naval Observatory, and other public and private donors. AMI is
operated by the University of Cambridge and supported by STFC. This
research has made use of NASA's Astrophysics Data System.

Facilities: \facility{AMI}, \facility{ATA}, \facility{Fermi},
\facility{INTEGRAL}, \facility{RATAN-600}, \facility{RXTE},
\facility{Swift}.

\end{document}